\documentclass[12pt,preprint]{aastex}
\usepackage{cite}
\usepackage{graphicx}
\usepackage{geometry}
\usepackage[nooneline,flushleft]{caption2}
\usepackage{natbib}
\usepackage{hyperref}  

\shorttitle{u-band Variables in LCSSPA}
\shortauthors{Cao et al.}
\begin{document}
\title{Spectral Identification of the u-band Variable Sources in Two LAMOST fields}
\author{Tian-Wen Cao, Ming Yang, Hong Wu, Tian-Meng Zhang, Jian-Rong Shi, Hao-Tong Zhang, Fan Yang, Jing-Kun Zhao, Xu Zhou, Zhou Fan, Zhao-Ji Jiang, Jun Ma, Jia-Li Wang, Zhen-Yu Wu, Hu Zou, Zhi-Min Zhou, Jun-Dan Nie,  A-Li Luo, Xue-Bing Wu, Yong-Heng Zhao }

\altaffiltext{1}{Key Laboratory of Optical Astronomy, National Astronomical Observatories, Chinese Academy of Sciences, Beijing 100012, P.R. China}

\begin{abstract}
We selected 82 u-band variable objects based on the u-band photometry data from SCUSS and SDSS, in the field of LAMOST Complete Spectroscopic Survey of Pointing Area at Southern Galactic Cap. The magnitude variation of the targets is restricted to larger than 0.2 mag and limiting magnitude down to 19.0 mag in u-band. According to the spectra from LAMOST, there are 11 quasars with redshift between 0.4 and 1.8, 60 variable stars and 11 unidentified targets. The variable stars include one active M-dwarf with a series of Balmer emission lines, seven Horizontal Branch stars containing six RR Lyrae stars matching with SIMBAD, and one giant, one AGB star and two RR Lyrae candidates by different color selections. All these variable stars mainly locate near the main sequence in the $g-r$ vs. $u-g$ diagram. The quasars are well distinguished from stars by both $u-g$ color and variation in u-band.

\end{abstract}

\keywords{quasars: general --- stars: general --- stars: statistics}

\section{Introduction \label{intro}}

Recent large surveys have provided astronomers an unprecedented opportunity in the field of the time-domain astronomy. The two current large projects of Pan-STARRS\citep{2002SPIE.4836..154K} and The Palomar Transient Factory( PTF; \citealt{2009PASP..121.1395L}), which cover large field of sky, have an ability to search many classes of variable objects with different timescales. The future Large Synoptic Survey Telescope (LSST; \citealt{2008arXiv0805.2366I, 2009arXiv0912.0201L}) can futher increase the number of variable objects in several order of magnitudes.
Though the photometry of multiple observations with different timescales can identify many variable objects, especially periodic variable stars, such as RR Lyrae, eclipsing binaries, and Delta Scuti, etc, spectral identification is still an important tool to verify some other types of variable objects, for example, quasars.

Some works have been done before on the Stripe 82 \citep{2007AJ....134..973I,2008MNRAS.386..887B,2010ApJS..186..233B}, which is an important region of the Sloan Digital Sky Survey (SDSS； \citealt{2000AJ....120.1579Y}) covered about 290-$degree^2$ on the celestial equator in the Southern Galactic Hemisphere. This region has been observed 10 times on average and many variable sources such as quasars, RR Lyrae stars, giant stars, eclipsing binaries have been confirmed by both photometry and spectroscopy.

Generally, quasars and blue stars are observed more effectively in u-band than in other optical bands, since their luminosities or variations in u-band are physically larger than those in other optical bands \citep{2012ApJ...748...58D}. Based on these features, quasars and some variable stars (such as RR Lyrae stars) are easily distinguished from others in the color-color diagram of $g-r$ vs. $u-g$  \citep{2002AJ....123.2945R,2011ApJ...728...26M}.

The South Galactic Cap u-band Sky Survey (SCUSS) \citep{2015AJ....150..104Z} is a project which finally performs sky survey of about 5000-$degree^2$ of the South Galactic Cap in u-band by using 2.3m Bok Telescope and the area is also almost covered by SDSS. Taking advantage of the u-band variation between SDSS and SCUSS, many variable objects (ex. quasars， \citealt{2015PASP..127...94Z}) can be selected. However the large follow-up spectral identifications should only depend on the large spectral survey programs.

The Large Sky Area Multi-Object Fiber Spectroscopic Telescope (LAMOST, also known as Guoshoujing telescope, GSJT, \citealt{1996ApOpt..35.5155W, 2004ChJAA...4....1S, 2012RAA....12.1197C, 2012RAA....12..723Z}) is a Wang-Su Schmidt telescope located in Xinglong Station of National Astronomical Observatory, Chinese Academy of Sciences (NAOC). LAMOST is a special reflecting Schmidt telescope with the correcting mirror Ma and the primary mirror Mb. An innovative active optics technique is utilized by LAMOST and it can achieve a series different reflecting Schmidt systems by changing its mirror surface continuously. With 4-meter clear aperture, 20-degree$^2$ field of view (FOV) and 4000 fibers, it is able to simultaneously spectroscopically observe more than 3000 scientific targets in a single exposure. This highly efficient facility will survey a large volume of space (including SCUSS area) for both stars and galaxies. By taking advantage of large area and multi-object fiber of LAMOST, the spectra of u-band variables can be acquired efficiently.

The LAMOST Complete Spectroscopic Survey of Pointing Area (LCSSPA) at Southern Galactic Cap (SGC) is a LAMOST Key Project， which designed to spectroscopically observe all sources (Galactic and extra-galactic) with limiting magnitude of $r=18.1~mag$ by selecting two 20-$degree^2$ FOVs at the SGC. The survey mainly aims at completeness of the LAMOST Extra GAlactic Survey (LEGAS). Meanwhile, besides of the normal galaxy survey, many scientific research fields are also included, such as the studies of galaxy clusters, luminous infrared galaxies, and  time-series variable sources (u-band variables and quasars). Thus the spectral identification of u-band variable sources of LCSSPA can be treated as a precursor research in future large program.

In this paper, we identify the u-band variables by their spectra in two fields of LCSSPA. The sample selection, observation and data reduction are presented in \textsection 2. The spectral identification are described in \textsection 3. In \textsection 4, we analyze the color and spectral characteristics of u-band variables, and discuss the reliability and efficiency of quasar selection. The summary is given in \textsection 5.

\section{Sample Selection, Observation and Data Reduction \label{samXobsXredu}}

The LCSSPA covers two 20-$degrees^2$ areas with central coordinates at $R.A. = 37.88^{\circ}$, $Dec. = 3.44^{\circ}$ ( Field A) and $R.A. = 21.53^{\circ}$, $Dec. = -2.20^{\circ}$ (Field B). These two areas are  in the overlapping field of SCUSS and SDSS. Thus, we are able to take spectroscopic observation of selected u-band variables in two LCSSPA fields. The targets in the FOV are mainly constituted of stars, galaxies, quasars, u-band variables and HII regions. According to their r-band magnitude, all targets in the FOV are divided into two kinds of plate, bright (B) plate, ($14.0 \sim 16.0mag$) and faint (F) plate, ($16.0 \sim 18.1mag$). The observations have been conducted from Sep 2012 to Jan 2014. The Field A was observed with 5 B and 12 F plates, and Field B was observed with 6 B and 5 F plates. The raw data have been reduced with LAMOST 2D and 1D pipelines \citep{2012RAA....12.1243L, 2014IAUS..298..428L} which including bias subtraction, flat-fielding through twilight exposures, cosmic-ray removal, spectrum extraction, wavelength calibration, sky subtraction and exposure coaddition.

The u-band limiting magnitude of the SCUSS is $\sim$23.2 mag which is about 1.5 mag deeper than that of SDSS. Based on the performance of LAMOST, we have restricted u-band photometry PSF magnitude down to 19.0 mag. Then, the coordinates of all targets in the FOV are matched between the SDSS and SCUSS data within one arcsecond. The final error ($\sigma$) of the magnitude variation in u-band combines both errors of SCUSS and SDSS, and the fitting of the 3-$\sigma$ errors of all targets is shown as red dashed lies in Figure.1. Here, a criterion of 0.2 magnituse is adopted to select the u-band variable objects.

\begin{figure}
\begin{center}
\resizebox{15cm}{!}{\includegraphics{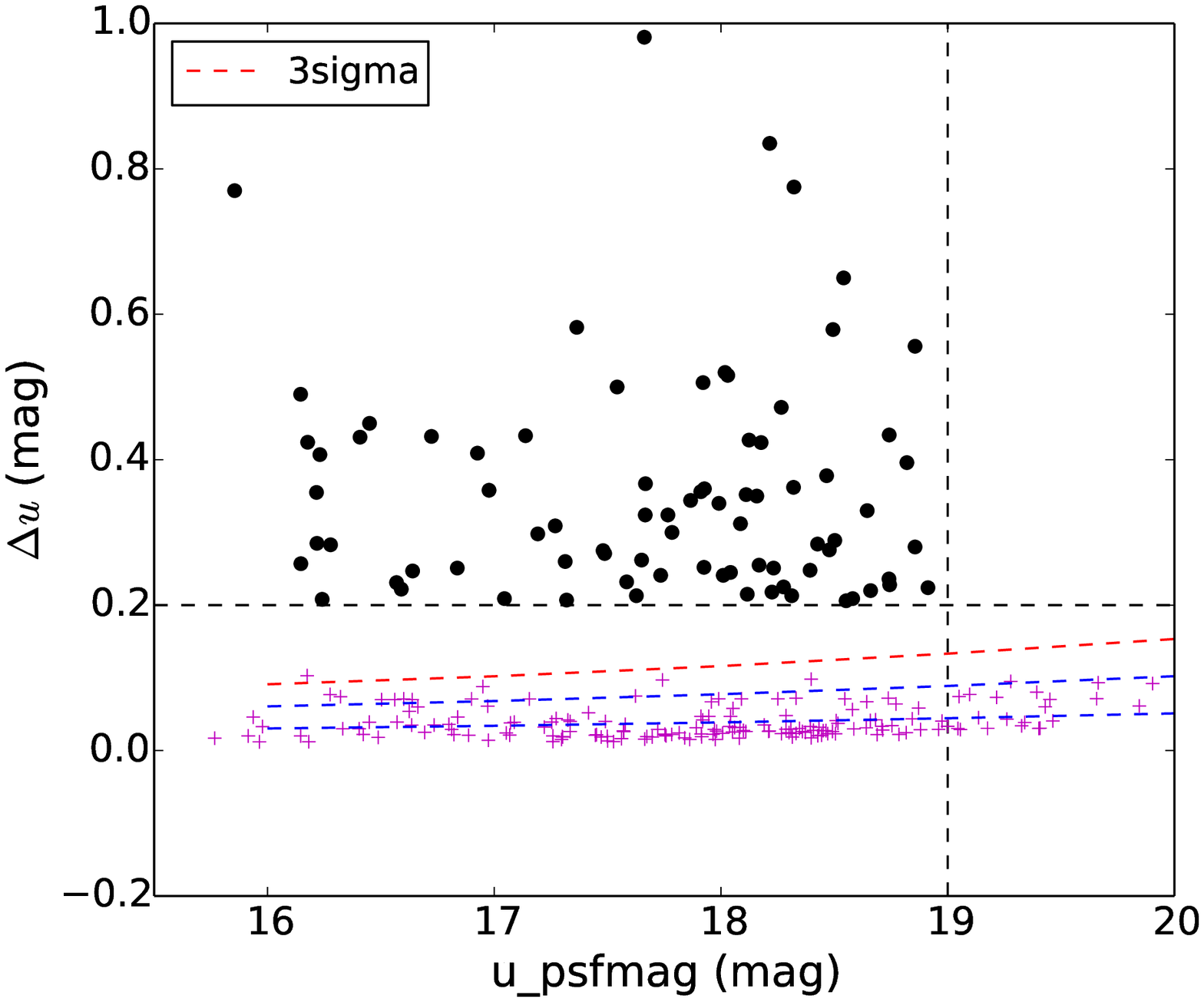}}
\end{center}
\caption{Diagram of magnitudes vesus variation in u-band. The plus signs (+) represent the standard deviation ($\sigma$) of magnitude varation. The red dashed line represents the fitting of 3-$\sigma$ errors by second order polynomial. The black dots represent  82 u-band variable objects. The dashed lines are selection criteria: u$<$ 19 and $\Delta$u $>$ 0.2. \label{devia}}
\end{figure}

 All the objects have been inspected via SDSS images and flags to eliminate those with bad photometry. The targets with any nearby objects less than 2-arcsec are eliminated too. Finally, There are 82 u-band variable sources are observed by LAMOST (see Figure\ref{devia}).

\section{Identification of u-band Variable Objects \label{Ident}}

\subsection{Quasar \label{quasar}}

Quasars are identified by their broad band emission lines: such as [C\uppercase\expandafter{\romannumeral4}] $\lambda$1549, [C\uppercase\expandafter{\romannumeral3}] $\lambda$1909, Mg\uppercase\expandafter{\romannumeral2} $\lambda$2800, H$\gamma$ $\lambda$4340, H$\beta$ $\lambda$4861, H$\alpha$ $\lambda$6563. There is a total of 11 quasars out of 82 variable sources. The redshifts are measured manually on LAMOST spectra with at least two emission lines, including above broad emission lines and some narrow emission lines such as [O\uppercase\expandafter{\romannumeral3}] $\lambda\lambda$4959,5007. The errors of redshifts are less than 0.001. The redshifts of quasars are between 0.4 and 1.8. All the information of quasars are listed in Table 2. Four quasars have been found in NED, among which two have also been observed by SDSS. The rest of seven quasars are newly discovered. The Figure\ref{quasar} shows the spectra of 11 quasars.

\subsection{Variable stars \label{variable}}

60 u-band variable stars have been identified by their optical spectra. The rest of 11 sources are unidentified due to the low signal-to-noise rate (SNR) of spectra. Figures 3-4 show all the spectra of these stars.

The MILES library \citep{2006MNRAS.371..703S, 2010MNRAS.404.1639V} which contains 985 high SNR low-resolution observed stellar spectra, provides important parameters of different stellar subtypes. MILES can be used as a source of templates to cassify stellar spectra \citep{2011RAA....11..563Z}. We have used the MILES library as the spectral template to cross-match with the spectra of 60 variable stars and determine their spectral types. The template matching is mainly based on the typical spectral profile and absorption lines. There are 5 A-type, 12 F-type, 32 G-type, 6 K-type and 5 M-type stars have been identified according to the template matching.

Meanwhile, the 60 variable stars are also cross-matched with LAMOST Date Release Two (DR2) which including the stellar parameters of effective temperature, surface gravity and [Fe/H ] derived by the LASP pipeline \citep{2014arXiv1407.1980W}. There are 51 objects matched and the distribution of spectral types is similar to the one derived from the use of MILES templates. Also, we have identified five M-dwarf including one active M-dwarf {\bf J022727.49+031054.8} with strong Balmer emission series (H8, H$\varepsilon$, H$\delta$, H$\gamma$, H$\beta$, H$\alpha$) and [Ca\uppercase\expandafter{\romannumeral2}] HK lines by spectral inspection.

After spectral classification, we cross-matched all these variable sources with SIMBAD \citep{2000A&AS..143....9W} except for quasars. There are seven Horizontal Branch stars including six RR Lyrae stars and one variable star candidate. And on the basis of color selection of RR Lyrae \citep{2005AJ....129.1096I} and other variable stars \citep{2009AJ....137.4377Y, 2010ApJS..186..233B}, we also selected one AGB, one giant and other two RR Lyrae star candidates. Table \ref{tbl-1} shows the criteria of color-color selection of stars. Table \ref{tbl-2} lists the general properties of all u-band variables including their spectral types.

\section{Discussion \label{dis}}

\subsection{Optical Color Diagrams \label{color}}

Due to warning flags in some bands of SDSS photometric data, only 39 stars, 11 quasars and 9 unidentified sources can be shown in the $g-r$ vs. $u-g$ diagram in Figure\ref{color-color}. From the diagram, the positions of variable sources  show that variable stars are located near the main sequence and quasars appear at the region
consistent with previous works on Stripe 82 \citep{2007AJ....134.2236S}. 
The variable sources which show up near the main sequence may be some eclipsing binary or some activity outbreaks (e.g. a magnetic storm in an M-type dwarf). For the unidentified sources, there is a need for follow-up observations in future studies.

In the Figure, we can see that the characteristic color of quasars is $u-g\leq0.6$ which is consistent with low-redshift quasars in SDSS \citep{2003MmSAI..74..978I}. Combining with $u-g$ color criteria and u-band magnitude variation $\geq0.2$ mag, quasars can be almost completed identified. Of course, some variables will be missed by setting the u-band magnitude variation $\geq0.2$ mag.

\subsection{Reliability of Variation \label{varia}}

Since the u-band filter of SCUSS is slightly bluer than SDSS and the flux calibration of SCUSS-u did not consider the color term, there exists a systematic magnitude difference between SCUSS and SDSS in u-band and the difference depends on $u-g$ color of targets. Based on the Equation 3 of \citet{2015AJ....150..104Z}, the systematic differences of all the variable stars in this paper are less than 0.026 mag with $u-g$ values from 0.8 to 2.5. If we adopt the Equation 1 of \citet{2015PASP..127...94Z}, the systematic differences are less than 0.036 mag. Therefore the systematic error of variable stars is less than 0.04 mag.

Meanwhile the magnitude difference of quasars changes with the redshift. \citet{2015PASP..127...94Z} showed the changes with redshift in their Figure 5.
The maximum of magnitude difference of composite quasar spectra is about 0.06 mag at redshift of 1.5 for low-redshift quasars($z < 2$) and less than 0.03 mag at redshift of $z<1.5$. Even considering some special quasars, the magnitude difference is less than 0.12 mag. As our quasars  are all low-redshift ($z < 2$), only two of them have redshifts larger than 1.5. However the two u-band magnitude variations are 0.232 and 0.252 mag respectively, their variations are still much higher than the possible systematic differences.

Because the SCUSS is about 1.5 magnitudes deeper than SDSS in u-band, the photometric error of SCUSS are far below that of SDSS. The final error of magnidtue variation is dominated by SDSS. Generally, most sample variable sources have errors less than 0.05 mag, and only a few of them could reach 0.08 mag.
Considering both systematic error and measurement error, it is quite safe to set a criteria of 0.2 mag to select variable sources in u-band. The variations are more than  three times of measurement errors, and also much higher than systematic errors even considering the worst case.

\subsection{The Spectra of Variable Stars \label{spec}}

Most absorption lines of stars appear at the spectral wavelengths in the LAMOST blue arm, so the SNR of blue arm is more important. However only 22 stars have  SNRs of higher than $20$ in blue arm, we mainly focus on the analysis of the absorption lines in these stars.

There are four RRLyr stars and one of them have three spectra obtained by different nights. The RRLyr star\citep{1996A&A...315..475c} is a kind of pulsational variable star whose luminosity, color and radial velocity periodically vary during the pulsation. The figure\ref{RRL} shows the H$\gamma$, H$\beta$ and H$\alpha$ profile of {\bf J013016.71-024240.2} in different day ( possibly different phases). In H$\alpha$ spectra, it also the presents the obviously fill-in H$\alpha$ emission line \citep{2014NewA...26...72Y} in two nights. The H$\alpha$ emission relates with the heating of shock wave at different phase\citep{2014A&A...565A..73G}.

Additional 10 variable stars' spectra are late F-type or early G-type, .shown in Figure\ref{fgspec} Their spectra show [Ca\uppercase\expandafter{\romannumeral2}]HK lines and G-band, which is caused by molecule CH(from $\lambda$4295 to $\lambda$4315), and weak balmer series features and the metallic lines. Those features are similar to the typical spectrum of G0V\citep{1978rmsa.book.....M}.

The other 8 variable stars' spectra are late G-type and K-type star. The G-type stars show some lines  sensetive with luminosity such as the G-band of CH and Sr\uppercase\expandafter{\romannumeral2}($\lambda$4077)\citep{1976aasc.book.....K}. However few of our spectra stars show strong Sr\uppercase\expandafter{\romannumeral2} feature of G-type. Furthermore the MgH($\lambda$4780)  can clearly be seen late-K-type\citep{1987clst.book.....J} stars. Figure\ref{gk} show expamples of a G-type star with Sr\uppercase\expandafter{\romannumeral2} feature and a late K-type star.

Based on the steller parameters acquired from the LAMOST DR2, most of variables are main sequence stars and this is consistent with the result in the color-color diagram. 

\subsection{ The Method and Efficiency of Quasar Selection\label{MEQS}}

\citet{2007AJ....134.2236S} employed the multi-color method  to select the variable sources in Stripe 82. More than 95\% of them are low-redshift quasars, RR Lyrae stars and main sequence stars. Due to the field of stripe 82 has multi-time observations, Bhatti\citep{2010ApJS..186..233B} has utilised the light-curve to study the variable sources and these variable sources include quasars and periodic variables(e.g. eclipsing binary systems, RR Lyrae and Delta Scuti candidates). \citet{0004-637X-714-2-1194} parameterized the single-band variability by a model of the light-curve to select quasars at $2.5<z<3.0$ with high efficiency. \citet{2015ApJ...811...95P} combines color and variability information to select quasars and the method effictivly improves the selection of the quasars at $2.7<z<3.5$. Though Pan-STARRS1(PS1) survey provides multi-color(grizy) observation, it is not quite efficiency to select quasars in lower redshift without u-band. However the varibility from multi-epoch observations will greatly increase the efficiency\citep{2014ApJ...784...92M}.

In all, combining with both multi-color and variability will provide a high efficiency method to select quasars. However we only have two-epoch with gap of four to twelve years (Table 2) and we can only obtain the variation instead of variability of sources. Fortunately, most quasar present the variation in timescale of years\citep{2007AJ....134.2236S} and u-band is very sensitive to the variation of quasar at lower redshift. With criteria of $u-g\leq0.6$ and magnitudes difference $\geq0.2$ mag in u-band, we identified 11 quasars from 12 candidates by LAMOST spectra, with an efficiency of 91.7\%. At high galactic latitude of LCSSPA, sources with $u-g\leq0.6$ are mainly quasars and white dwarfs. Because the variation of white dwarfs are too small to reach 0.2 mag, we believe the left one is still a real quasar.
To explian the critical role of variation in u-band, we compare the efficiency by only $u-g$ selection at the same field. With $u-g\leq0.6$, we have 155 sources with LAMOST spectra. Except the 59 unidentified sources, we identified 48 quasars. The efficiency to select quasars is only about 30\%.
Though the sample may be not enough to give the true efficientcy of the method, it is a quite efficient way. However, we missed most quasars in this field. The possible reason is that many quasars could have variation far less than 0.2 mag at the two-epoch in u-band.
The quasar selection with two-epoch variation more than three times the photometric errors \citep{2015PASP..127...94Z} in overlapping field of SCUSS and SDSS at the South Galactic Cap would overcome part of our uncompleteness.

In recent future, many multi-epoch survey programs would focus on variable sources. However, all these surveys (such as PS1, LSST, etc on) do not include u-band. Therefore, the u-band of both SCUSS and SDSS is an unique and important band to help us to separate the quasars ($z<2$) and variable stars by criteria of either color or variation.

\section{Summary \label{sum}}

In this paper, we  selected a sample of 82 u-band variables in the region of LCSSPA based on criteria of u-band magnitude variation $\geq0.2$ mag and the SCUSS u-band magnitude brighter than 19.0 mag. After spectral identification, there are 11 quasars with redshift between 0.4 and 1.8, 60 variable stars and 11 unidentified sources. Among the variable stars, there is one active M-dwarf, which present a series of strong Balmer emission lines; There are seven HB star including six RR Lyrae stars and one variable star candidate matching with SIMBAD; one AGB, one giant and two RR Lyrae candidates by multi-color diagnostic tools.
Basing on the analysis of the reliability and efficiency, we conclude that the criteria of both $u-g\leq0.6$ and u-band variation above 0.2 mag is quite an efficiency to select quasar at redshift less than 2.

\section*{Acknowledgments}

This project is supported by the China Ministry of Science and Technology under the State Key Development Program for Basic Research (2014CB845705, 2012CB821800); the National Natural Science Foundation of China (Grant No.11173030, 11225316); the Strategic Priority Research Program "The Emergence of Cosmological Structures" of the Chinese Academy of Sciences (Grant No.XDB09000000);
the Guoshoujing Telescope Spectroscopic Survey Key Projects; Wu,Z.Y. was supported by the Chinese National Natural Science Foundation (Grant Nos. 11373033).

Guoshoujing Telescope (the Large Sky Area Multi-Object Fiber Spectroscopic Telescope, LAMOST) is a National Major Scientific Project built by the Chinese Academy of Sciences.
Funding for the project has been provided by the National Development and Reform Commission. LAMOST is operated and managed by the National Astronomical Observatories, Chinese Academy of Sciences.

The SCUSS is funded by the Main Direction Program of Knowledge Innovation of Chinese Academy of Sciences (No. KJCX2-EW-T06). It is also an international cooperative project between National Astronomical Observatories, Chinese Academy of Sciences and Steward Observatory, University of Arizona, USA. Technical supports and observational assistances of the Bok telescope are provided by Steward Observatory. The project is managed by the National Astronomical Observatory of China and Shanghai Astronomical Observatory.

\clearpage
\bibliographystyle{apj}
\bibstyle{thesisstyle}
\bibliography{uband.bib}

\clearpage

\clearpage
\newgeometry{left=0.0cm,bottom=1cm}
\begin{figure}[h!tb]
\captionstyle{flushleft}
\onelinecaptionstrue
\begin{center}
\includegraphics[width=8in]{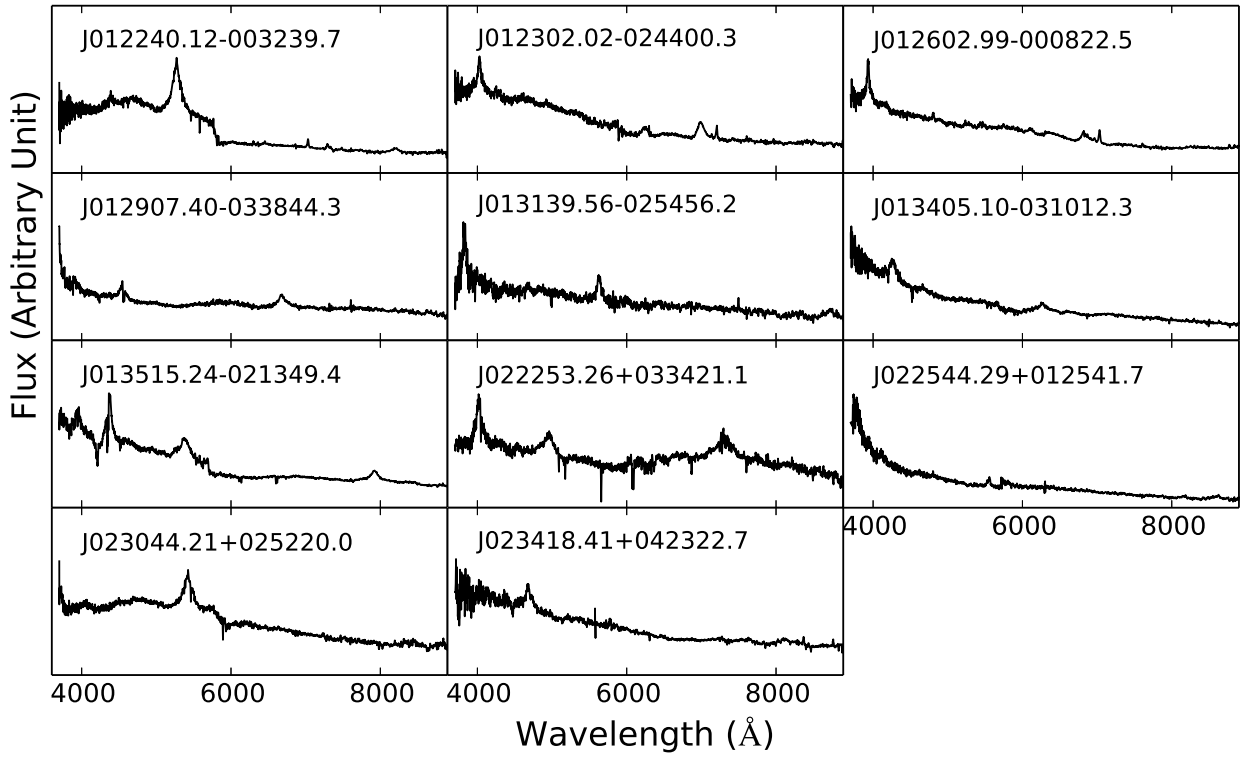}
\end{center}
\caption{The spectra of 11 variable quasars from LCSSPA.
       \label{quasar}}
\end{figure}

\clearpage
\newgeometry{left=0.0cm,bottom=1cm}
\begin{figure}[h!tb]
\captionstyle{flushleft}
\onelinecaptionstrue
\begin{center}
\includegraphics[width=8.5in]{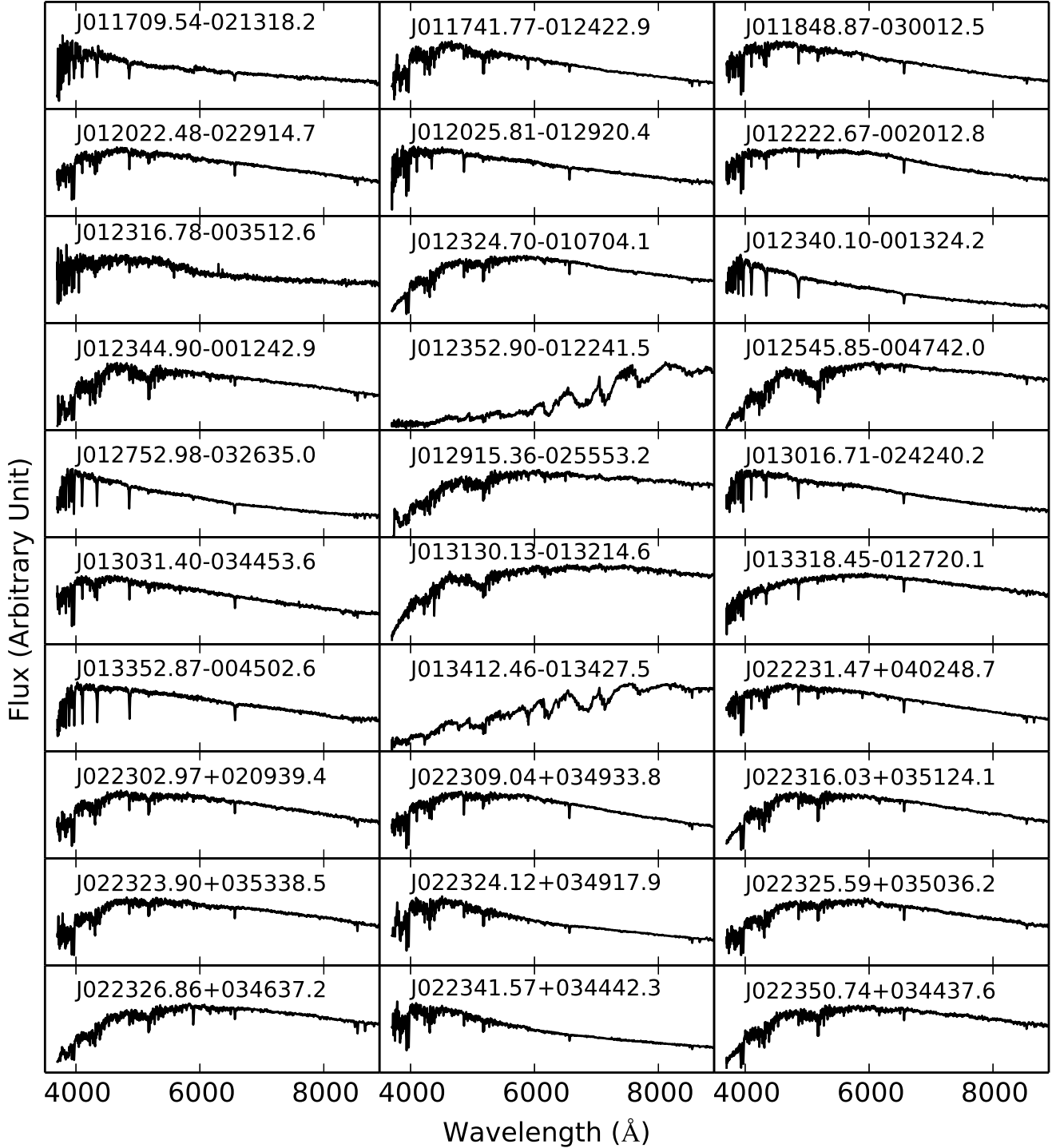}
\end{center}
\caption{The spectra of variable stars in u-band from LCSSPA.
       \label{spectra1}}
\end{figure}

\clearpage
\newgeometry{left=0.0cm,bottom=1cm}
\begin{figure}[h!tb]
\captionstyle{flushleft}
\onelinecaptionstrue
\begin{center}
\includegraphics[width=8.5in]{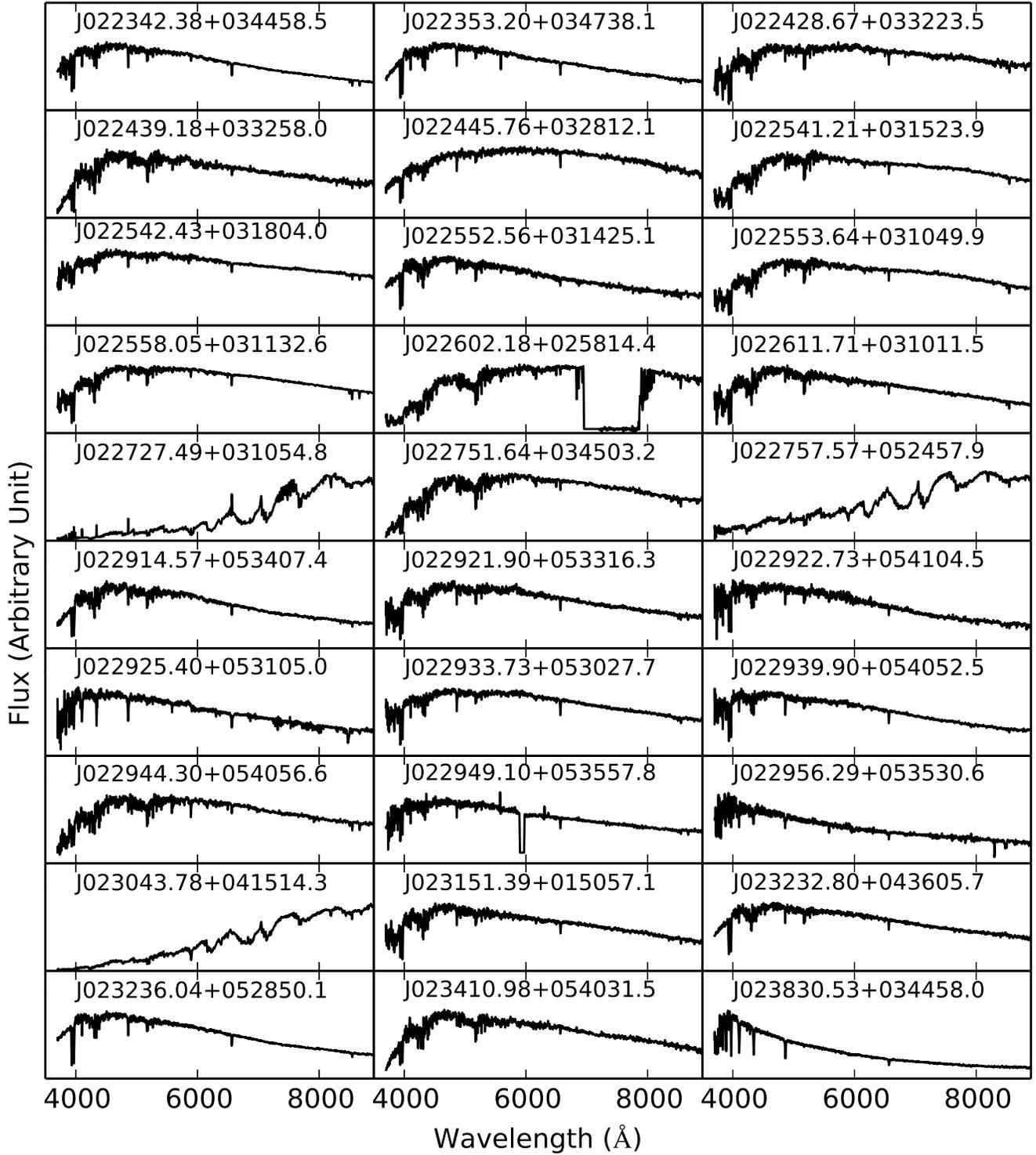}
\end{center}
\caption{The Same as Figure 3.
       \label{spectra2}}
\end{figure}

\clearpage
\restoregeometry
\begin{figure}[h!tb]
\begin{center}
\resizebox{15cm}{!}{\includegraphics{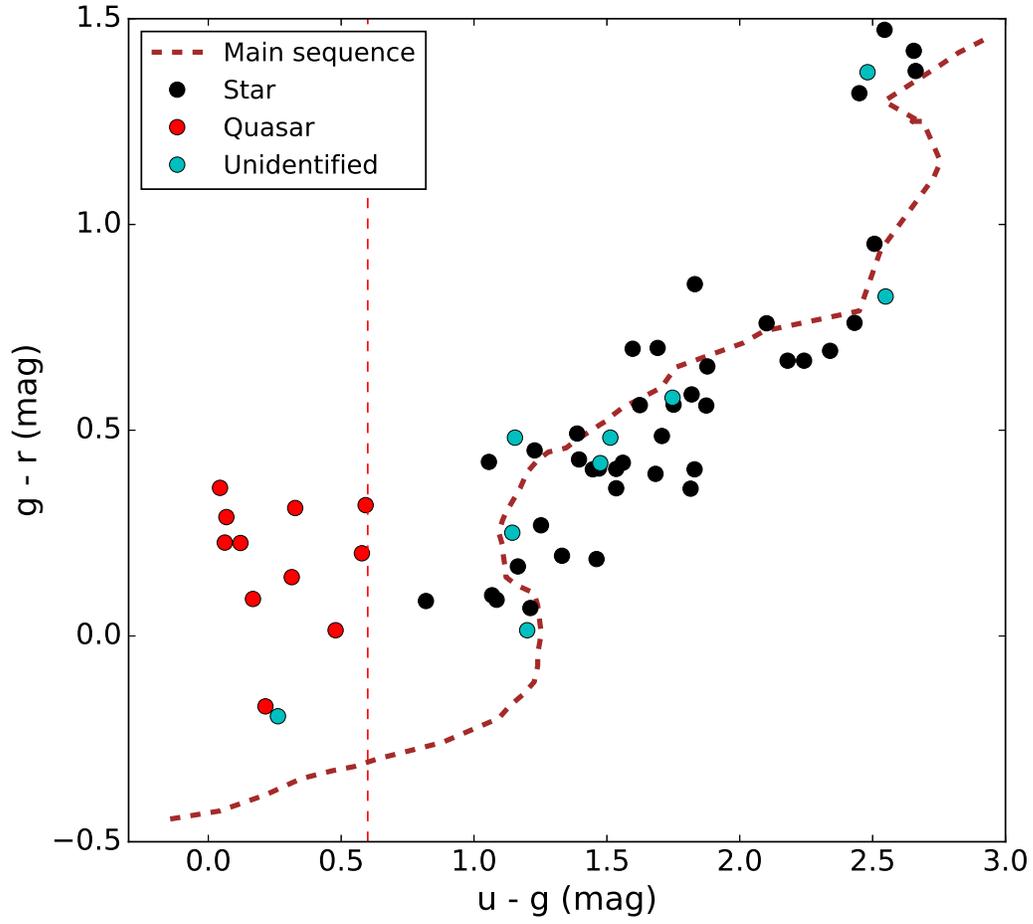}}
\end{center}
\caption{The $g-r$ vs. $u-g$ color-color diagram of u-band variable sources. The red, black and blue dots represent quasars, stars and unidentified sources, respectively (same below). The brown dashed line represents the position of main-sequence. The red dashed line represents the line of quasar criteria of $u-g=0.6$. \label{color-color}}
\end{figure}

\clearpage
\newgeometry{left=0.0cm,bottom=1cm}
\begin{figure}[h!tb]
\includegraphics[width=8.5in]{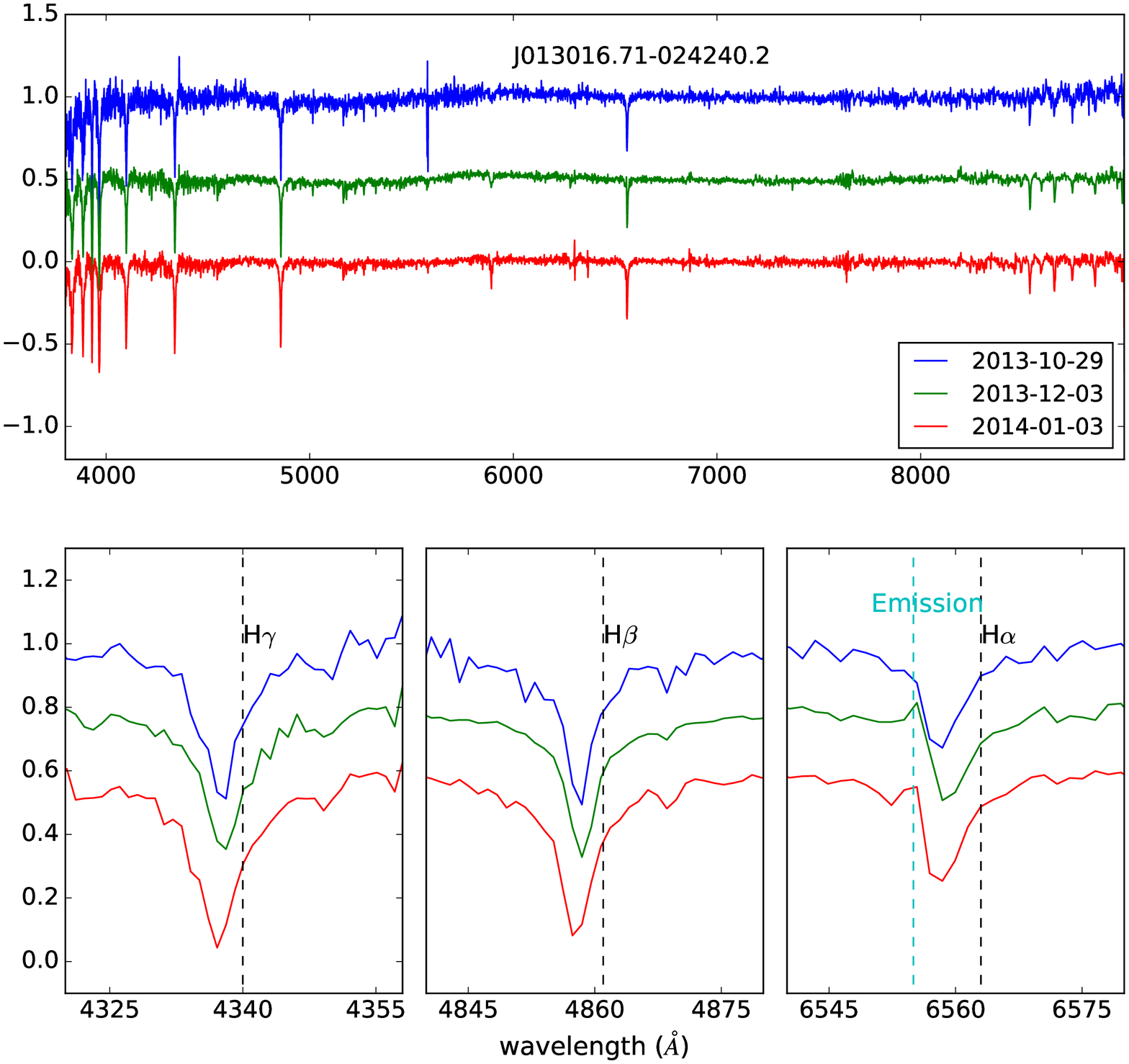}
\captionsetup{margin=70pt}
\caption{The top panel shows three normalized spectra at different nights. The bottom panels shows Balmer absorption lines and the vertical black dashed show the restframe wavelength of spectral lines. The vertical shallow blue dashed in the right picture shows the position of fill-in H$\alpha$ emission.
    \label{RRL}}
\end{figure}

\clearpage
\newgeometry{left=0.0cm,bottom=1cm}
\begin{figure}[h!tb]
\includegraphics[width=8.5in]{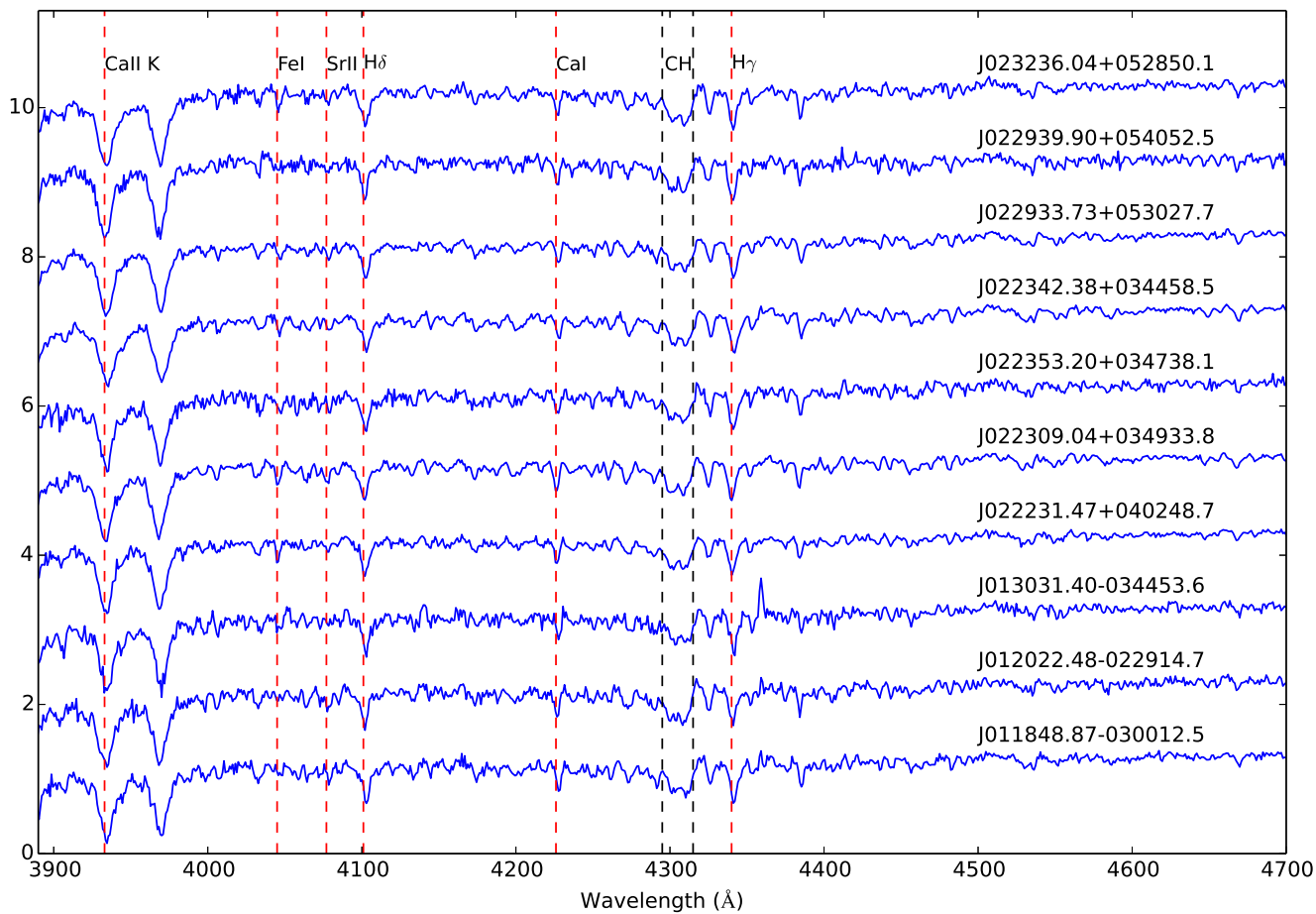}
\captionsetup{margin=70pt}
\caption{The spectra of 10 objects whose spectrum have similar features. The molecule CH(from $\lambda$4295 to $\lambda$4315) is wide, and shown between two black dashed lines. The red lines represent other the positions of other lines.
    \label{fgspec}}
\end{figure}

\clearpage
\newgeometry{left=0.0cm,bottom=1cm}
\begin{figure}[h!tb]
\includegraphics[width=9in]{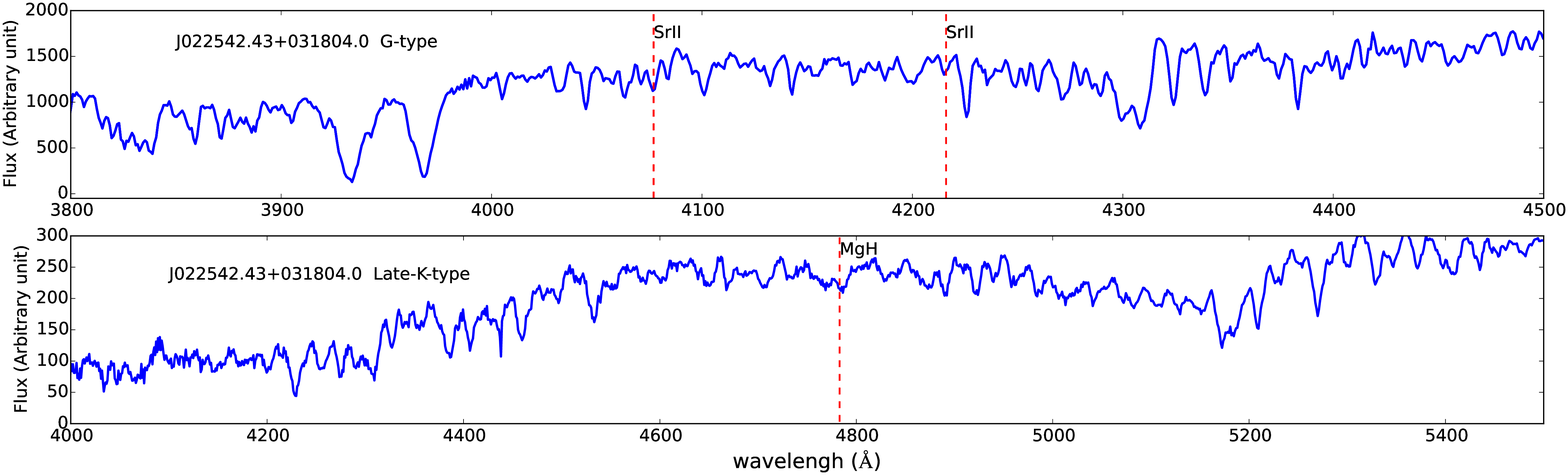}
\captionsetup{margin=70pt}
\caption{The example spectra of G-type and late-K-type stars. The upper is a G-type star with SrII features and the lower is a late K-type star with MgH feature.
    \label{gk}}
\end{figure}

\clearpage
\begin{deluxetable}{cc}
\tabletypesize{\scriptsize}
\tablecaption{Characteristic Color of Different Types of Stars \label{tbl-1}}
\tablewidth{0pt}
\tablehead{
\colhead{Target Type} & \colhead{{Color Selecion}\tablenotemark{a}}
}
\startdata
RR Lyrae candidate\tablenotemark{b} & $0.98\textless(u-g)\textless1.30$  and  $-0.05\textless RR1 \textless0.35$ \\
  & and $0.06\textless RR2\textless0.55$  and  $-0.15\textless(r-i)\textless0.22$\\
  & and  $-0.21\textless(i-z)\textless0.25$ \\
  &  \\
sdO/sdB/white dwarf\tablenotemark{c} & $-0.1\textless(g-r)\textless-0.2$  and  $-0.1\textless(u-g)\textless0.7$  \\
  & and  $u-g+2*(g-r)\textless-0.1$\\
  &  \\
AGB\tablenotemark{c} & $2.5\textless(u-g)\textless3.5$  and  $0.9\textless(g-r)\textless1.3$\\
  & and  $s\textless-0.06$ \\
  &  \\
k gaint\tablenotemark{c} & $0.7\textless(u-g)\textless4.0$  and  $0.35\textless(g-r)\textless0.7$  \\
  & and  $0.15\textless(r-i)\textless0.6$  and  $l\textgreater 0.07$\\
  & \\
Low metallicity\tablenotemark{c}  & $-0.5\textless(g-r)\textless0.75$  and  $0.6\textless(u-g)\textless3.0$\\
  & and $l\textgreater 0.135 $ \\
\enddata
\tablenotetext{a}{The color adopted in selection are defined below :\\
$\mathrm{RR1}=0.45*(u-g)-(g-r)-0.12$\\
$\mathrm{RR2}=(u-g)+0.67*(g-r)-1.07$ \\
$\mathrm{s}=-0.249*u+0.795*g-0.555*r+0.124$\\
$\mathrm{l}=-0.436*u+1.129*g-0.119*r-0.574*i+0.1984$\\}
\tablenotetext{b}{Using color from \citealt{2007AJ....134.2236S}.}
\tablenotetext{c}{Using color from \citealt{2009AJ....137.4377Y}.}
\end{deluxetable}

\clearpage
\begin{deluxetable}{ccccccccccccc}
\tabletypesize{\tiny}
\rotate
\tablecolumns{13}
\tablecaption{The General Properties of the u-band Variable objects \label{tbl-2}}
\tablewidth{0pt}
\tablehead{
\colhead{Name} &\colhead{RA ($^\circ$)} & \colhead{Dec ($^\circ$)} & \colhead{$\mathrm{u_{SCUSS}}$}  & \colhead{date (UTC)} & \colhead{$\mathrm{u_{SDSS}}$ } & \colhead{date (UT)} & \colhead{$\mathrm{g_{SDSS}}$} & \colhead{$\mathrm{r_{SDSS}}$}  & \colhead{Redshift} & \colhead{{Type}\tablenotemark{a}} &\colhead{$\mathrm{T_{eff}}$} &\colhead{$\mathrm{log_{g}}$}
}

\startdata
J023418.41+042322.7 & 38.576725  & 4.3896499  & 18.552 & 2012-12-09 & 18.346 & 2008-10-03 & 17.867 & 17.853 & 0.671 &quasar &---- &----\\
 & & &$\pm$0.009 & &$\pm$0.029 & &$\pm$0.025 &$\pm$0.025 &$\pm$0.001 & & & \\
J023044.21+025220.0 & 37.6842194 & 2.8722501  & 18.042 & 2013-12-01 & 18.287 & 2008-09-06 & 17.695 & 17.377 & 0.937 &quasar &---- &----\\
 & & &$\pm$0.004 & &$\pm$0.029 & &$\pm$0.029 &$\pm$0.021 &$\pm$0.001 & & & \\
J022544.29+012541.7 & 36.4345703 & 1.428275   & 18.108 & 2013-01-09 & 18.444 & 2008-10-03 & 18.323  & 18.097 & 0.996 &quasar &---- &----\\
 & & &$\pm$0.007 & &$\pm$0.020 & &$\pm$0.025 &$\pm$0.039 &$\pm$0.004 & & & \\
J022253.26+033421.1 & 35.7219238 & 3.5725334  & 17.925 & 2013-12-26 & 17.673 & 2009-09-16 & 17.505  & 17.415 & 1.601 &quasar &---- &----\\
 & & &$\pm$0.003 & &$\pm$0.019 & &$\pm$0.025 &$\pm$0.016 &$\pm$0.003 & & & \\
J013515.24-021349.4 & 23.8135242 & -2.2304027 & 17.584 & 2013-10-25 & 17.816 & 2008-10-31 & 17.754 & 17.527 & 1.818 &quasar*&---- &----\\
 & & &$\pm$0.004 & &$\pm$0.024 & &$\pm$0.024 &$\pm$0.016 &$\pm$0.002 & & & \\
J013405.10-031012.3 & 23.521265  & -3.1700835 & 17.866 & 2013-11-07 & 18.210 & 2008-10-31 & 18.142  & 17.853 & 1.235 &quasar &---- &----\\
 & & &$\pm$0.005 & &$\pm$0.026 & &$\pm$0.033 &$\pm$0.016 &$\pm$0.003 & & & \\
J013139.56-025456.2 & 22.914854  & -2.9156139 & 18.177 & 2013-10-25 & 18.413 & 2008-10-30 & 18.086 & 17.775 & 1.008 &quasar &---- &----\\
 & & &$\pm$0.006 & &$\pm$0.105 & &$\pm$0.019 &$\pm$0.024 &$\pm$0.003 & & & \\
J012907.40-033844.3 & 22.2808456 & -3.6456554 & 17.911 & 2013-11-03 & 18.267 & 2008-10-31 & 18.223  & 17.863 & 1.385 &quasar &---- &----\\
 & & &$\pm$0.004 & &$\pm$0.023 & &$\pm$0.028 &$\pm$0.019 &$\pm$0.002 & & & \\
J012602.99-000822.5 & 21.5124702 & -0.1396056 & 17.921 & 2013-11-03 & 18.427  & 2001-11-18 & 18.113  & 17.97 & 0.402 &quasar*&---- &----\\
 & & &$\pm$0.004 & &$\pm$0.020 & &$\pm$0.025 &$\pm$0.010 &$\pm$0.001 & & & \\
J012302.02-024400.3 & 20.7584324 & -2.7334332 & 17.627 & 2013-12-05 & 17.840 & 2009-01-16 & 17.625 & 17.796 & 0.439 &quasar*&---- &----\\
 & & &$\pm$0.004 & &$\pm$0.017 & &$\pm$0.024 &$\pm$0.020 &$\pm$0.001 & & & \\
J012240.12-003239.7 & 20.6671867 & -0.5443611 & 18.426& 2013-12-05 & 18.710 & 2004-09-24 & 18.132 & 17.931 & 0.886 &quasar*&---- &----\\
 & & &$\pm$0.005 & &$\pm$0.033 & &$\pm$0.021 &$\pm$0.018 &$\pm$0.003 & & & \\
J023830.53+034458.0 & 39.6272469 & 3.7494667 & 18.494 & 2013-10-02 & 17.915 & 2008-10-08 & 16.750 & 16.581 & ---- & A7IV/F0/RRLyr & 5812.05 & 3.657\\
 & & &$\pm$0.006 & &$\pm$0.018 & &$\pm$0.020 &$\pm$0.018 & & &$\pm$449.91 &$\pm$1.26 \\
J023410.98+054031.5 & 38.5457878 & 5.6754303 & 18.478 & 2013-12-26 & 18.754 & 2005-10-12 & 16.653 & 15.893 & ---- & G5/G8/-- & 4991.17& 4.538\\
 & & &$\pm$0.005 & &$\pm$0.034 & &$\pm$0.026 &$\pm$0.020 & & &$\pm$207.94 &$\pm$0.403 \\
J023236.04+052850.1 & 38.1500549 & 5.4804583 & 15.855 & 2014-01-05 & 16.332 & 2009-09-16 & 15.092 & 14.37 & ---- & G0/G0/--  &5882.35 & 4.368 \\
 & & &$\pm$0.002 & &$\pm$0.030 & &$\pm$0.024 &$\pm$0.034 & & &$\pm$255.62 &$\pm$0.563 \\
J023232.80+043605.7 & 38.1366959 & 4.6015997 & 16.640 & 2014-01-05 & 16.887 & 2008-10-03 & 15.263 & 14.702 & ---- & G3/G3/--  &5441.25 & 3.883\\
 & & &$\pm$0.002 & &$\pm$0.021 & &$\pm$0.028 &$\pm$0.029 & & &$\pm$263.21 &$\pm$0.741 \\
J023151.39+015057.1 & 37.9641304 & 1.8492193 & 16.147 & 2013-12-01 & 16.404 & 2008-10-03 & 15.117 & 14.737 & ---- &F9/F9/--  &5746.46  & 4.491\\
 & & &$\pm$0.002 & &$\pm$0.03  & &$\pm$0.022 &$\pm$0.06 & & &$\pm$260.2 &$\pm$0.536 \\
J023043.78+041514.3 & 37.6824493 & 4.2539721 & 18.741 & 2013-12-01 & 18.505 & 2008-10-08 & 15.960 & 14.487 & ---- & M/--/dwarf &---- &----\\
 & & &$\pm$0.007 & &$\pm$0.022 & &$\pm$0.028 &$\pm$0.014 & & & & \\
J022956.29+053530.6 & 37.4845581 & 5.5918474 & 18.856 & 2012-10-13 & 19.136 & 2004-12-14 & 17.805 & 17.61 & ---- & A/--/-- &---- &----\\
 & & &$\pm$0.005 & &$\pm$0.045 & &$\pm$0.027 &$\pm$0.025 & & & & \\
J022949.10+053557.8 & 37.4546127 & 5.5993972 & 18.158& 2013-01-06 & 18.508 & 2004-12-14 & 16.973 & 16.614& ---- & F7/F7/-- &5924.64  &4.202\\
 & & &$\pm$0.004 & &$\pm$0.041 & &$\pm$0.027 &$\pm$0.025 & & &$\pm$312.95 &$\pm$0.565 \\
J022944.30+054056.6 & 37.4346161 & 5.6823969 & 18.466 & 2013-01-06 & 18.844 & 2004-12-14 & 16.412 & 15.651 & ---- & G8/K0/-- &5128.96  &4.531\\
 & & &$\pm$0.008 & &$\pm$0.043 & &$\pm$0.027 &$\pm$0.025 & & &$\pm$140.81 &$\pm$0.366 \\
J022939.90+054052.5 & 37.4162865 & 5.681272 & 16.216 & 2013-12-01 & 16.571 & 2004-12-14 & 15.011 & 14.59 & ---- & F6/F2/-- &5971.51  &4.197\\
 & & &$\pm$0.003 & &$\pm$0.039 & &$\pm$0.027 &$\pm$0.025 & & &$\pm$256.22 &$\pm$0.559 \\
J022933.73+053027.7 & 37.3905525 & 5.5077028 & 16.977 & 2013-12-01 & 17.335& 2004-12-14 & 15.800 & 15.394 & ---- & G1/F7/-- &5901.83  &4.194\\
 & & &$\pm$0.003 & &$\pm$0.040 & &$\pm$0.027 &$\pm$0.025 & & &$\pm$252.71 &$\pm$0.56 \\
J022925.40+053105.0 & 37.3558655 & 5.5180745 & 17.192 & 2012-11-13 & 17.490 & 2004-12-14 & 16.029 & 15.842 & ---- & F5/F0/-- &6975.91  &4.262\\
 & & &$\pm$0.003 & &$\pm$0.040 & &$\pm$0.027 &$\pm$0.025 & & &$\pm$309.96 &$\pm$0.38 \\
J022922.73+054104.5 & 37.3447151 & 5.6845999 & 17.045 & 2013-10-30 & 17.254 & 2004-12-14 & 15.859 & 15.430 & ---- & F9/G0/-- &5657.46  &3.757 \\
 & & &$\pm$0.003 & &$\pm$0.039 & &$\pm$0.025 &$\pm$0.024 & & &$\pm$396.87 &$\pm$0.797 \\
J022921.90+053316.3 & 37.3412819 & 5.5545392 & 16.837 & 2013-12-01 & 17.088 & 2004-12-14 & 15.381 & 14.895 & ---- & G4/G7/-- &5628.7    &4.131\\
 & & &$\pm$0.002 & &$\pm$0.039 & &$\pm$0.025 &$\pm$0.024 & & &$\pm$275.8  &$\pm$0.57 \\
J022914.57+053407.4 & 37.3107262 & 5.5687275 & 16.241 & 2014-01-05 & 16.449 & 2004-12-14 & 14.631 & 14.115 & ---- & G3/G5/-- &5585.42  & 4.153\\
 & & &$\pm$0.002 & &$\pm$0.039 & &$\pm$0.025 &$\pm$0.024 & & &$\pm$189.57 &$\pm$0.546 \\
J022757.57+052457.9 & 36.9898796 & 5.4160862 & 18.312 & 2013-12-01 & 18.099 & 2009-09-16 & 15.437 & 14.064 & ---- & M/--/M dwarf &---- &----\\
 & & &$\pm$0.006 & &$\pm$0.027 & &$\pm$0.023 &$\pm$0.021 & & & & \\
J022751.64+034503.2 & 36.9651985 & 3.7509055 & 17.138 & 2013-12-26 & 17.571 & 2009-09-16 & 15.409 & 14.389 & ---- & K0/K1/-- &5070.75 &4.599  \\
 & & &$\pm$0.002 & &$\pm$0.026 & &$\pm$0.023 &$\pm$0.023 & & &$\pm$79.48 &$\pm$0.32 \\
J022727.49+031054.8 & 36.8645668 & 3.1819 & 18.018 & 2013-10-02 & 17.498 & 2008-09-06 & 15.159 & 14.053 & ---- & M /--/M dwarf+ &---- &----\\
 & & &$\pm$0.004 & &$\pm$0.019 & &$\pm$0.015 &$\pm$0.001 & & & & \\
J022611.71+031011.5 & 36.5487976 & 3.1698694 & 17.542 & 2013-12-01 & 18.042 & 2009-09-16 & 15.862 & 15.193 & ---- & G7/G8/-- &5068.55 &3.801\\
 & & &$\pm$0.003 & &$\pm$0.047 & &$\pm$0.024 &$\pm$0.022 & & &$\pm$214.5 &$\pm$0.574 \\
J022602.18+025814.4 & 36.5090904 & 2.9706779 & 18.393 & 2013-10-02 & 18.641 & 2008-10-03 & 16.134 & 15.181 & ---- & G7/K5/AGB* &4596.76 &4.715\\
 & & &$\pm$0.004 & &$\pm$0.031 & &$\pm$0.016 &$\pm$0.02 & & &$\pm$90.82 &$\pm$0.252 \\
J022558.05+031132.6 & 36.4918976 & 3.1923974 & 16.408 & 2013-01-09 & 16.839 & 2009-09-16 & 14.850 & 14.194 & ---- & G8/G5/-- &5108.47 &3.466 \\
 & & &$\pm$0.002 & &$\pm$0.046 & &$\pm$0.024 &$\pm$0.001& & &$\pm$98.81 &$\pm$0.625 \\
J022553.64+031049.9 & 36.4735222 & 3.1805501 & 17.650 & 2013-10-02 & 17.912 & 2009-09-16 & 16.093 & 15.506 & ---- & G8/G7/--  &5358.32 &4.528\\
 & & &$\pm$0.003 & &$\pm$0.047 & &$\pm$0.024 &$\pm$0.022 & & &$\pm$176.85 &$\pm$0.486 \\
J022552.56+031425.1 & 36.4690056 & 3.2403224 & 17.927 & 2013-12-26 & 18.287 & 2009-09-16 & 16.552 & 16.013 & ---- & G2/G3/-- &5274.4   &4.406 \\
 & & &$\pm$0.003 & &$\pm$0.048 & &$\pm$0.024 &$\pm$0.022 & & &$\pm$228.15 &$\pm$0.51 \\
J022542.43+031804.0 & 36.4268227 & 3.3011334 & 17.784 & 2013-01-09 & 18.084 & 2009-09-16 & 16.614 & 16.207 & ---- & G3/F9/-- &5942.6   &4.631\\
 & & &$\pm$0.003 & &$\pm$0.032 & &$\pm$0.014 &$\pm$0.015 & & &$\pm$252.22 &$\pm$0.441 \\
J022541.21+031523.9 & 36.4217224 & 3.2566583 & 17.364 & 2013-10-02 & 17.946 & 2009-09-16 & 15.606 & 14.913 & ---- & G9/G8/-- &5343.25  &4.688\\
 & & &$\pm$0.003 & &$\pm$0.047 & &$\pm$0.024 &$\pm$0.022 & & &$\pm$134.38 &$\pm$0.406 \\
J022445.76+032812.1 & 36.1906891 & 3.4700305 & 17.766 & 2013-12-26 & 18.090 & 2009-09-16 & 16.558 & 16.115 & ---- & G0/G2/-- &5904.52   &4.55\\
 & & &$\pm$0.003 & &$\pm$0.071 & &$\pm$0.022 &$\pm$0.009 & & &$\pm$413.6 &$\pm$0.931 \\
J022439.18+033258.0 & 36.1632729 & 3.5494637 & 18.820 & 2013-12-26 & 19.216 & 2009-09-16 & 16.854 & 16.127 & ---- & G7/G7/-- &5110.81  &4.559\\
 & & &$\pm$0.006 & &$\pm$0.073 & &$\pm$0.022 &$\pm$0.009 & & &$\pm$231.75 &$\pm$0.418 \\
J022428.67+033223.5 & 36.1194649 & 3.5398805 & 18.030 & 2013-10-02 & 18.546 & 2009-09-16 & 16.672 & 16.112 & ---- & G8/G7/-- &5417.57  &4.769\\
 & & &$\pm$0.005 & &$\pm$0.071 & &$\pm$0.022 &$\pm$0.009 & & &$\pm$228.35 &$\pm$0.478 \\
J022353.20+034738.1 & 35.9716988 & 3.7939305 & 17.734 & 2013-12-26 & 17.975 & 2009-09-16 & 16.528 & 16.123 & ---- & G2/F7/-- &5895.04 & 4.203 \\
 & & &$\pm$0.003 & &$\pm$0.015 & &$\pm$0.012 &$\pm$0.014 & & &$\pm$263.23 &$\pm$0.64 \\
J022350.74+034437.6 & 35.9614182 & 3.7438056 & 18.266 & 2013-12-26 & 18.738 & 2009-09-16 & 16.496 & 15.827 & ---- & K1/G9/-- &5119.65  &4.522\\
 & & &$\pm$0.004 & &$\pm$0.072 & &$\pm$0.016 &$\pm$0.025 & & &$\pm$161.25 &$\pm$0.431 \\
J022342.38+034458.5 & 35.9265823 & 3.7495973 & 16.177& 2013-01-05 & 16.601 & 2009-09-16 & 14.898 & 16.887 & ---- & G2/G0/-- &6090.33  &4.37 \\
 & & &$\pm$0.002 & &$\pm$0.071 & &$\pm$0.015 &$\pm$0.032 & & &$\pm$151.68 &$\pm$0.489 \\
J022341.57+034442.3 & 35.923233 & 3.7450917 & 16.723 & 2012-11-14 & 17.155 & 2009-09-16 & 15.271 & 14.070 & ---- & F6/F9/-- &5593.81  &4.479\\
 & & &$\pm$0.002 & &$\pm$0.071 & &$\pm$0.016 &$\pm$0.001 & & &$\pm$171.41 &$\pm$0.509 \\
J022326.86+034637.2 & 35.8619308 & 3.7770195 & 17.666 & 2013-01-06 & 17.990 & 2009-09-16 & 15.841 & 15.224 & ---- & K2/K0/-- &5279.59  &4.425 \\
 & & &$\pm$0.003 & &$\pm$0.071 & &$\pm$0.016 &$\pm$0.002 & & &$\pm$126.79 &$\pm$0.434 \\
J022325.59+035036.2 & 35.8566437 & 3.8433971 & 17.991 & 2013-12-01 & 18.331& 2009-09-16 & 16.580 & 16.018 & ---- & G8/G5/-- &5610.74  &4.649\\
 & & &$\pm$0.003 & &$\pm$0.072 & &$\pm$0.016 &$\pm$0.025 & & &$\pm$180.87 &$\pm$0.486 \\
J022324.12+034917.9 & 35.8505325 & 3.8216584 & 16.231 & 2012-11-14 & 16.638 & 2009-09-16 & 14.667 & 14.059 & ---- & F8/F9/-- &5416.3   &4.46\\
 & & &$\pm$0.002 & &$\pm$0.07  & &$\pm$0.015 &$\pm$0.001 & & &$\pm$121.23 &$\pm$0.478 \\
J022323.90+035338.5 & 35.8495827 & 3.8940361 & 18.086 & 2013-10-30 & 18.398 & 2009-09-16 & 16.520 & 15.865 & ---- & G8/G7/-- &5221.7   &3.666\\
 & & &$\pm$0.004 & &$\pm$0.017 & &$\pm$0.012 &$\pm$0.014 & & &$\pm$243.34 &$\pm$0.7 \\
J022316.03+035124.1 & 35.8168106 & 3.8567026 & 16.450 & 2013-12-26 & 16.900 & 2009-09-16 & 14.782 & 14.116 & ---- & G8/G7/-- &5207.6   &4.616 \\
 & & &$\pm$0.002 & &$\pm$0.071 & &$\pm$0.015 &$\pm$0.001 & & &$\pm$112.99 &$\pm$0.42 \\
J022309.04+034933.8 & 35.7876816 & 3.8260748 & 16.278 & 2013-12-01 & 16.561 & 2009-09-16 & 15.040 & 17.014 & ---- & G4/G2/-- &5977.43  &4.355 \\
 & & &$\pm$0.002 & &$\pm$0.070 & &$\pm$0.016 &$\pm$0.056 & & &$\pm$149.89 &$\pm$0.507 \\
J022302.97+020939.4 & 35.762394 & 2.1609502 & 16.569 & 2013-12-01 & 16.800 & 2008-10-03 & 15.459& 15.023 & ---- & G6/G3/-- &5573.42  &4.524\\
 & & &$\pm$0.003 & &$\pm$0.036 & &$\pm$0.036 &$\pm$0.050 & & &$\pm$214.96 &$\pm$0.534 \\
J022231.47+040248.7 & 35.6311302 & 4.0468779 & 16.146 & 2013-01-08 & 16.636& 2009-09-16 & 14.953& 14.559 & ---- & G3/G0/-- &5867.2   &4.166 \\
 & & &$\pm$0.002 & &$\pm$0.035 & &$\pm$0.017 &$\pm$0.013 & & &$\pm$193.08 &$\pm$0.609 \\
J013412.46-013427.5 & 23.5519257 & -1.5743166 & 18.276 & 2013-12-03 & 18.051 & 2008-10-30 & 15.601 & 14.282 & ---- & M/--/M dwarf &---- &----\\
 & & &$\pm$0.005 & &$\pm$0.025 & &$\pm$0.014 &$\pm$0.028 & & & & \\
J013352.87-004502.6 & 23.470295 & -0.7507361 & 18.856 & 2013-12-05 & 18.300 & 2003-11-20 & 17.215 & 17.127 & ---- & A8III/F0/RRLyr &6215.14&4.597\\
 & & &$\pm$0.009 & &$\pm$0.023 & &$\pm$0.019 &$\pm$0.013 & & &$\pm$419.92 &$\pm$0.417 \\
J013318.45-012720.1 & 23.3268909 & -1.4555889 & 18.124 & 2013-11-03 & 17.697 & 2008-10-31 & 16.485 & 16.417 & ---- & F/F0/RRLyr &6174.39  &4.164\\
 & & &$\pm$0.005 & &$\pm$0.018 & &$\pm$0.016 &$\pm$0.013 & & &$\pm$305.13 &$\pm$0.511 \\
J013130.13-013214.6 & 22.8755569 & -1.5374084 & 18.502 & 2014-01-03 & 18.213 & 2008-10-31 & 15.820 & 14.779 & ---- & K7/K5/V* &4313.16 &4.358 \\
 & & &$\pm$0.007 & &$\pm$0.026 & &$\pm$0.024 &$\pm$0.019 & & &$\pm$85.22 &$\pm$0.279 \\
J013031.40-034453.6 & 22.6308327 & -3.748225 & 18.742 & 2013-11-03 & 18.308 & 2008-12-30 & 17.252 & 16.829 & ---- & F9/F8/-- &6015.54  &4.351\\
 & & &$\pm$0.005 & &$\pm$0.029 & &$\pm$0.03 &$\pm$0.016 & & &$\pm$235.89 &$\pm$0.473 \\
J013016.71-024240.2 & 22.5696545 & -2.7111723 & 16.269 & 2013-10-29 & 15.915 & 2008-10-31 & 14.756 &14.611 & ---- & F0/F0/ RRLyr&5995.88&4.021\\
 & & &$\pm$0.002 & &$\pm$0.020 & &$\pm$0.021 &$\pm$0.001 & & &$\pm$269.78 &$\pm$0.568 \\
J012915.36-025553.2 & 22.3140125 & -2.9314556 & 17.319 & 2013-12-03 & 17.526 & 2008-10-30 & 15.695 & 14.840 & ---- & K3/K3/-- &4864.59 &3.944\\
 & & &$\pm$0.003 & &$\pm$0.012 & &$\pm$0.024 &$\pm$0.018 & & &$\pm$79.36 &$\pm$0.368 \\
J012752.98-032635.0 & 21.9707565 & -3.443064 & 18.116 & 2013-10-25 & 18.331 & 2008-10-31 & 17.079 & 16.810 & ---- & F5/F0/RRLyr &5836.9   &3.631\\
 & & &$\pm$0.005 & &$\pm$0.023 & &$\pm$0.025 &$\pm$0.034 & & &$\pm$314.49 &$\pm$0.831 \\
J012545.85-004742.0 & 21.4410629 & -0.7950194 & 17.662 & 2013-10-29 & 18.643 & 2003-11-20 & 15.499 & 14.551 & ---- & K0/K5/-- & 4523.2 &4.629\\
 & & &$\pm$0.003 & &$\pm$0.067 & &$\pm$0.027 &$\pm$0.019 & & &$\pm$67.22 &$\pm$0.233 \\
J012352.90-012241.5 & 20.9704208 & -1.3782084 & 18.581 & 2013-12-27 & 18.372& 2008-10-31 & 15.717 & 14.295 & ---- & M/--/M dwarf &---- &---- \\
 & & &$\pm$0.007 & &$\pm$0.024 & &$\pm$0.014 &$\pm$0.012 & & & & \\
J012344.90-001242.9 & 20.9371166 & -0.2119194 & 17.313 & 2013-10-29 & 17.053 & 2003-11-20 & 15.362&14.662&----& G8/K0/gaint* &5020.74 &4.609\\
 & & &$\pm$0.003 & &$\pm$0.024 & &$\pm$0.022 &$\pm$0.023 & & &$\pm$106.14 &$\pm$0.344 \\
J012340.10-001324.2 & 20.917099 & -0.2234056 & 18.232 & 2013-11-03 & 17.981 & 2003-11-20 & 17.162 & 17.077 & ---- & A/--/HB* &---- &----\\
 & & &$\pm$0.006 & &$\pm$0.026 & &$\pm$0.022 &$\pm$0.023 & & & & \\
J012324.70-010704.1 & 20.8529301 & -1.1178167 & 16.590 & 2014-01-09 & 16.812 & 2003-11-20 & 15.424 & 14.932 & ---- & G9/G8/-- &5424.32 &4.617 \\
 & & &$\pm$0.002 & &$\pm$0.029 & &$\pm$0.021 &$\pm$0.013 & & &$\pm$168.26 &$\pm$0.468 \\
J012316.78-003512.6 & 20.8199348 & -0.5868361 & 18.322 & 2013-12-05 & 19.097 & 2001-09-18 & 17.267 & 16.862 & ---- & G/--/-- &---- &----\\
 & & &$\pm$0.005 & &$\pm$0.077 & &$\pm$0.021 &$\pm$0.015 & & & & \\
J012222.67-002012.8 & 20.5944881 & -0.3368917 & 18.541 & 2013-11-03 & 17.891 & 2003-10-19 & 16.877 & 16.346 & ---- & G3/F0/-- &5587.11 &4.127\\
 & & &$\pm$0.005 & &$\pm$0.031 & &$\pm$0.024 &$\pm$0.022 & & &$\pm$208.86 &$\pm$0.603 \\
J012025.81-012920.4 & 20.1075497 & -1.4890195 & 18.215 & 2013-11-07 & 19.050 & 2008-10-31 & 17.235 & 16.877 & ---- & F5/F0/-- &6175.75& 4.204 \\
 & & &$\pm$0.006 & &$\pm$0.074 & &$\pm$0.024 &$\pm$0.012 & & &$\pm$270.61 &$\pm$0.462 \\
J012022.48-022914.7 & 20.0936794 & -2.4874222 & 18.225 & 2013-11-07 & 18.007 & 2008-10-30 & 16.779 &16.328 & ---- & G3/G0/-- &5809.82 & 4.047 \\
 & & &$\pm$0.006 & &$\pm$0.023 & &$\pm$0.017 &$\pm$0.015 & & &$\pm$235.0 &$\pm$0.643 \\
J011848.87-030012.5 & 19.7036362 & -3.0034778 & 17.480 & 2013-11-07 & 17.755 & 2001-11-11 & 16.257 & 15.829 & ---- & G1/F9/-- &5840.47& 4.268 \\
 & & &$\pm$0.004 & &$\pm$0.021 & &$\pm$0.021 &$\pm$0.01 & & &$\pm$200.63 &$\pm$0.549 \\
J011741.77-012422.9 & 19.4240532 & -1.4063694 & 18.111 & 2013-10-25 & 18.463 & 2008-10-31 & 16.866 & 16.168 & ---- & G2/G7/--&5163.18  &4.5 \\
 & & &$\pm$0.005 & &$\pm$0.027 & &$\pm$0.018 &$\pm$0.018 & & &$\pm$132.06 &$\pm$0.45 \\
J011709.54-021318.2 & 19.2897625 & -2.2217417 & 18.66 & 2013-11-07 & 18.880 & 2008-10-31 & 17.812 & 17.713 & ---- & A1IV/--/RRLyr* &---- &----\\
 & & &$\pm$0.01 & &$\pm$0.027 & &$\pm$0.014 &$\pm$0.016 & & & & \\
J023831.91+030933.6 & 39.632981 & 3.1593502 & 18.168 & 2013-01-06 & 17.913 & 2008-09-06 & 17.651 & 17.846 & ---- & unidentified &---- &----\\
 & & &$\pm$0.006 & &$\pm$0.022 & &$\pm$0.017 &$\pm$0.018 & & & & \\
J023815.35+030800.7 & 39.563972 & 3.1335295 & 18.913 & 2013-10-30 & 18.689 & 2008-09-06 & 16.208 & 14.838 & ---- & unidentified &---- &----\\
 & & &$\pm$0.009 & &$\pm$0.02 & &$\pm$0.014 &$\pm$0.014 & & & & \\
J022944.83+053721.0 & 37.436792 & 5.6225182 & 18.320 & 2012-10-13 & 18.682 & 2004-12-14 & 17.207 & 16.787 & ---- & unidentified &---- &----\\
 & & &$\pm$0.005 & &$\pm$0.042 & &$\pm$0.027 &$\pm$0.025 & & & & \\
J022803.74+014932.9 & 37.015621 & 1.8258154 & 17.667 & 2013-10-30 & 17.300 & 2008-10-03 & 16.100 & 16.086 & ---- & unidentified/RRLyr* &---- &----\\
 & & &$\pm$0.003 & &$\pm$0.019 & &$\pm$0.018 &$\pm$0.015 & & & & \\
J022748.45+042834.5 & 36.951904 & 4.4762608 & 18.744 & 2013-12-26 & 18.516 & 2008-10-03 & 17.801 & 17.472 & ---- & unidentified &---- &----\\
 & & &$\pm$0.005 & &$\pm$0.037 & &$\pm$0.017 &$\pm$0.024 & & & & \\
J022447.56+032608.7 & 36.198182 & 3.4357571 & 16.218 & 2014-01-05 & 16.503 & 2009-09-16 & 15.024 & 15.298 & ---- & unidentified &---- &----\\
 & & &$\pm$0.002 & &$\pm$0.070 & &$\pm$0.022 &$\pm$0.002 & & & & \\
J022420.03+033027.8 & 36.083472 & 3.5077244 & 18.010 & 2014-01-05 & 18.251 & 2009-09-16 & 15.702 & 14.877 & ---- & unidentified &---- &----\\
 & & &$\pm$0.005 & &$\pm$0.071 & &$\pm$0.022 &$\pm$0.009 & & & & \\
J022219.15+040233.3 & 35.57979 & 4.0425862 & 17.269 & 2013-12-01 & 17.578 & 2009-09-16 & 16.065 & 15.583 & ---- & unidentified &---- &----\\
 & & &$\pm$0.003 & &$\pm$0.036 & &$\pm$0.017 &$\pm$0.014 & & & & \\
J022213.77+040535.8 & 35.557383 & 4.0933007 & 18.645 & 2013-12-26 & 18.975 & 2009-09-16 & 17.228 & 16.649 & ---- & unidentified &---- &----\\
 & & &$\pm$0.006 & &$\pm$0.040 & &$\pm$0.017 &$\pm$0.014 & & & & \\
J013156.26-023100.7 & 22.984421 & -2.5168671 & 17.487 & 2013-11-07 & 17.758 & 2008-10-30 & 16.604 & 16.122 & ---- & unidentified &---- &----\\
 & & &$\pm$0.003 & &$\pm$0.020 & &$\pm$0.016 &$\pm$0.03 & & & & \\
J011742.01-020819.8 & 19.425079 & -2.1388566 & 16.926 & 2013-11-07 & 17.335 & 2001-11-11 & 16.191 & 15.94 & ---- & unidentified/RRLyr &---- &----\\
 & & &$\pm$0.003 & &$\pm$0.026 & &$\pm$0.022 &$\pm$0.018 & & & & \\
\enddata
\tablenotetext{a}{For quasars, the ones which have been found in NED or SDSS are marked with asterisk. For stars, there are three types are marked in table including: (1) spectral types cross-matched with MILES; (2) spectral types cross-matched with LAMOST DR1; (3) stellar types identified by SIMBAD, color selection (asterisk) or spectral inspection (plus). Due to the low SNR, 11 variable sources can not be identified by MILES or LAMOST DR1.}

\end{deluxetable}
\clearpage

\end{document}